\documentclass[sigconf]{acmart}
\usepackage{booktabs} 
\usepackage{url}
\usepackage{microtype}
\usepackage{subfigure}
\usepackage{bm,amsmath}
\usepackage{paralist}
\usepackage{adjustbox}
\usepackage{multirow}
\usepackage{graphicx, graphics}
\usepackage{color, colortbl}
\definecolor{Gray}{gray}{0.9}

\setcopyright{rightsretained}

\newcommand{\figref}[1]{Figure \ref{#1}}

\newcommand{\argmax}
{\mathop{\mathrm{argmax}}}

\acmYear{2018}
\copyrightyear{2018}

\begin{document}
\title{A Structure-Oriented Unsupervised Crawling Strategy for Social Media Sites}

\author{Keyang Xu}
\affiliation{%
  \institution{Language Technologies Institute}
  \institution{Carnegie Mellon University}
  \city{Pittsburgh} 
  \state{PA}
  \country{USA}
  \postcode{15213}
}
\email{xky0714@gmail.com}

\author{Kyle Yingkai Gao}
\affiliation{%
  \institution{Language Technologies Institute}
  \institution{Carnegie Mellon University}
  \city{Pittsburgh} 
  \state{PA}
  \country{USA}
  \postcode{15213}
}
\email{kyle.ygao@gmail.com}

\author{Jamie Callan}
\affiliation{%
  \institution{Language Technologies Institute}
  \institution{Carnegie Mellon University}
  \city{Pittsburgh} 
  \state{PA} 
  \country{USA}
  \postcode{15213}
}
\email{callan@cs.cmu.edu}



\begin{abstract}
Existing techniques for efficiently crawling social media sites rely on URL patterns, query logs, and human supervision. This paper describes \texttt{SOUrCe}, a structure-oriented unsupervised crawler that uses page structures to learn how to crawl a social media site efficiently. \texttt{SOUrCe} consists of two stages. During its unsupervised learning phase, \texttt{SOUrCe} constructs a sitemap that clusters pages based on their structural similarity and generates a navigation table that describes how the different types of pages in the site are linked together. During its harvesting phase, it uses the navigation table and a crawling policy to guide the choice of which links to crawl next.  Experiments show that this architecture supports different styles of crawling efficiently, and does a better job of staying focused on user-created contents than baseline methods. 
\end{abstract}

\keywords{Web Crawling; Page Clustering; Link Prediction; }

\maketitle
\section{Introduction}
\label{sec:intro}

Web crawlers use a variety of methods to identify and download a useful set of web pages within time and bandwidth constraints.  There are several differences between social media crawlers and general web crawlers. One important difference is the opportunity for a social media crawler to use information about how a site is organized to crawl it more efficiently. Prior research observed that social media sites tend to have large regions of consistent structures where web pages are generated dynamically using a small number of page templates and content stored in a database. Knowing this structure enables a crawler to prioritize some types of pages (e.g., recent user-generated content) over others, or to spread its effort evenly to obtain a representative sample~\cite{DBLP:conf/www/CaiYLWZ08,DBLP:conf/vldb/ChoS07,jiang2013focus}, or to avoid downloading the same page via different URLs~\cite{sigir17-KeyangXu}.
Information about how the site is organized is provided manually, recognized by heuristics, or learned by recognizing consistent patterns in the site~\cite{DBLP:conf/sigir/VidalSMC06,DBLP:conf/www/CaiYLWZ08,zhang2008profile,DBLP:conf/webi/GuoLZZ06,jiang2013focus}. 

Most prior work uses URL patterns to represent page templates and to classify uncrawled links. However, URL naming conventions are not necessarily stable for pages created at different times, and crawlers must cope with duplication and redirection problems.~Simple patterns may be learned without supervision, but websites with complex URL patterns may require training data that must be generated by other methods~\cite{DBLP:conf/wsdm/KoppulaLACGS10,DBLP:conf/www/LeiCYKFZ10}. During the crawling phase, other resources, supervision, or policies are required to convert a set of URL patterns and a queue of uncrawled links into link scores or traversal paths that guide the crawler.

Using page templates to represent site structure is a powerful intuition. However the correspondence between URL patterns and page templates is a weak signal for fully exploiting this intuition.  This paper proposes a new approach to crawling social media that infers the presence of page templates from the structure of the web pages themselves, rather than from their URLs.  Web pages are longer and have more structure than URLs, which provides more information for inferring the presence of page templates and for deciphering how they are interconnected.  Our contributions are three-fold:
\begin{itemize}
\item An automatic method of clustering and classifying web pages based only on structural information;
\item A crawler that predicts the page type of an uncrawled link without the help of URL patterns, content information, or human supervision; and
\item A crawling framework that naturally supports several distinct crawling policies.
\end{itemize}
Experimental results show that the new approach to crawling social media is better than or comparable to existing methods that require external information.

The rest of this paper is organized as follows. Section \ref{sec:related} gives a brief review of related prior work. Sections \ref{sec:setting} and \ref{sec:architecture} describe the problem setting, framework, and \texttt{SOUrCe} crawler architecture.  Section \ref{sec:exp} reports on data collection, and experiments with \texttt{SOUrCe} and several baseline methods. Finally, Section \ref{sec:con} summarizes the paper and concludes. 

\section{Related Work}
\label{sec:related}
We review two lines of studies that are relevant to our work: web crawling strategy and web pages clustering. 

\subsection{Web Crawling Strategy}

Typically web crawlers traverse the web, which is a connected graph, by recursively recognizing links in a page to other pages, and downloading those other pages.  Due to resource limitations such as bandwidth, crawlers must be selective in choosing which among its discovered links to crawl next. A good intuition is weighting links based on their usefulness and importance, which can be estimated by PageRank and in-degree. However, link structure can be difficult to recognize accurately while crawling. Prior research estimates importance using partial  graphs \cite{cho1998efficient, DBLP:conf/vldb/ChoS07}, historically crawled data \cite{baeza2005crawling}, cash flow \cite{DBLP:conf/www/AbiteboulPC03} and other features such as the host domain and page titles \cite{alam2012novel}. 

Cho, et al.~\cite{cho1998efficient} proposed to weight links based on in-links count and partial PageRank calculation. Baeza-Yates, et al.~\cite{baeza2005crawling} indicated using historically crawled data is effective in ordering a crawling queue, as well as estimating the PageRank of unseen links. OPIC \cite{DBLP:conf/www/AbiteboulPC03} utilizes the concept of cash flow and estimates the importance of a link by ``cash'' that is distributed equally by pages that point to it. RankMass~\cite{DBLP:conf/vldb/ChoS07} uses all available links during crawling to calculate a lower bound on PageRank to prioritize the crawling queue. Fractional PageRank~\cite{alam2012novel} incorporates features such as host domain and page title in computing PageRank scores and proposes to reduce the computational cost by skipping the path towards a set of already downloaded pages. 

The recent importance of social media requires crawlers that are effective within specific sites. \texttt{iRobot} \cite{DBLP:conf/www/CaiYLWZ08} clusters pages using both repetitive regions and URL patterns, and filters out unimportant clusters via an informativeness measure. In its crawling phase, a traversal path is generated, which is a spanning tree on the graph, to guide crawlers. However, the process of generating a traversal path requires human supervision. Importance of clusters can be estimated by hub and authority scores~\cite{liu2011user}. However, this work utilizes search logs, which is an additional external resource. 

Vidal et al.\cite{DBLP:conf/sigir/VidalSMC06} propose \texttt{TPM}, which focuses on crawling a specific type of pages, known as ``target'', from one website. An example page is first used to find pages that share similar structure in a set of pages; then \texttt{TPM} learns URL navigation patterns that lead to target pages efficiently. Jiang et al.\cite{jiang2013focus} propose \texttt{FoCuS}, which is a supervised method using regular expressions to find thread and index pages in web forums. Zhang et al. \cite{zhang2008profile} explore focused crawling in social media by classifying pages into pre-defined groups. However, both methods are supervised \cite{jiang2013focus,zhang2008profile}, which can be not generalizable when crawling other sites. 
\subsection{Web Pages Clustering}
Web pages clustering methods can be categorized into four types based on features employed: link-based, content-based, URL-based and structure-based. 

Link-based methods \cite{cooley1997web,dean1999finding} state that links are considered to be topically similar if they share similar parent nodes. Content-based approaches \cite{hotho2002ontology,selamat2004web} categorize web pages according to the words and sentences contained. These methods argue that pages with similar functions share similar topic features.  The vector space model \cite{salton1988term} is commonly used to represent documents. 
However, content-based approaches are limited due to vagueness of topical differences and difficulties in understanding cross-language websites. URL-based methods~\cite{DBLP:conf/www/LeiCYKFZ10,DBLP:conf/wsdm/KoppulaLACGS10,hernandez2012statistical} have an advantage of clustering pages without downloading files. However, dealing with duplication~\cite{sigir17-KeyangXu} and redirection problems is a non-trival task. Once the patterns of URLs are decided, regular expression can be used to cluster and classify web pages. 
Structure-based approaches are more general ways to cluster web pages. Some previous research estimates the similarity of pairwise web pages through their tree structures~\cite{demaine2007optimal,nierman2002evaluating}. However, it is computationally expensive to do pairwise page comparison and not suitable for web-scale applications. Another idea is representing web pages with vector space models. For example, ~Crescenzi et al.~\cite{crescenzi2005clustering} indicate that two pages are considered to be structurally similar if they have common links placed in the same location. Minimum Description Length Principle (\texttt{MDL}) is adopted in combining small clusters. Grigalis et al \cite{grigalis2014using} further explore this idea by exploiting Xpaths in recording positions of specific links. Their method, known as UXCluster, is claimed to be more accurate and efficient than the method produced by pairwise distance.  
Cai et al \cite{DBLP:conf/www/CaiYLWZ08} propose to cluster web pages represented by repetitive region patterns extracted by web wrappers \cite{zhai2005web,zheng2007joint}. A drawback of most of structure-based clustering is that it requires heuristic threshold to decide whether to form a new cluster or to merge into an existing cluster. 

\section{Problem setting}
\label{sec:setting}
We begin by introducing some definitions and notations that are used in this paper. Then we present two important tasks: (1) Using layout structure to represent and cluster web pages; and (2) analyzing the potential destination cluster of a given anchor link.

\subsection{Definitioins}

\textbf{Page Type:} For a given social media site, pages that serve the same purpose are usually generated by the same layout template. Each layout template defines a distinct page type. We denote the collection of page types for a given site as $\mathcal{T}=\{T_1,T_2...T_m\}$ and $|\mathcal{T}|=m$. 

\textbf{Page Cluster:} We use a clustering algorithm to produce page clusters, each of which contains structurally similar pages. We denote the collection of page clusters as $\mathcal{C}=\{C_1,C_2...C_n\}$ and $|\mathcal{C}|=n$. The crawler's task is to produce a collection of clusters $\mathcal{C}$ that matches as closely as possible to the collection of (hidden) page types $\mathcal{T}$.

\textbf{Sitemap:} Following the previous definition~\cite{DBLP:conf/www/CaiYLWZ08}, a sitemap is a directed graph consisting of a set of vertices and existing arcs, where each vertex represents a page cluster and each arc indicates an observed anchor link. {Multiple arcs between a pair of clusters $(C_i, C_j)$ indicates that there are multiple links from pages in cluster $C_i$ to pages in cluster $C_j$.}

\textbf{Adjacency Matrix:} 
At the cluster levels, we define the adjacency matrix for graph $G=(V,E,\mathcal{C})$ as $A$, where $\mathcal{C}$ are the page clusters for set of pages $V$ and set of directed anchor links $E$. The size of $A$ is $|\mathcal{C}| \times |\mathcal{C}|$ and its element $A_{ij}$ represents the weight between the directed pair ($C_i$,$C_j$), which is the total number of links from any node $V_m \in C_i$ directed to $V_n \in C_j$. 

\subsection{Structural features for Web Pages}
\label{sec:structural_feature}

A web page can be represented as a tree structure, following the Document Object Model (DOM). Nodes in the hierarchical DOM tree represent HTML elements.  We use a bag-of-Xpaths model to represent the DOM tree. An Xpath is defined as a string that concatenates HTML elements from the root to a leaf node. We only extract Xpaths that end with a leaf node, such as anchor links, texts or images, because they are the elements displayed in browsers. Similarly to the bag-of-words model for text retrieval, we use term frequency and inverse document frequency~\cite{salton1988term} to weight an Xpath $x$ in a web page $d$ as follows: 
\begin{align}
w_{x,d} =\log ({tf_{x,d}+1}) \times \log\left({\frac{|D|}{{df}_{x,D}} + 1}\right)
\end{align}
${tf}_{x,d}$ is the frequency of Xpath $x$ in document $d$, $|D|$ is the size of page collections $D$, and $df_{x,D}$ is the number of pages that contain Xpath $x$ in $D$. The log of ${tf}_{x,d}$ prevents high frequency Xpaths from dominating the clustering process. Inverse document frequency reduces the weight of Xpaths that appear on almost every page.  After extracting features for each page, we apply L1 normalization:
\begin{align}
w_{x,d} = \frac{w_{x,d}}{\sum_{x'}{w_{x',d}}}
\end{align}
Euclidean distance is used to measure the similarity between two pages.

\subsection{Analyzing Anchor Links }

\begin{figure}[t]
\includegraphics[width=\linewidth]{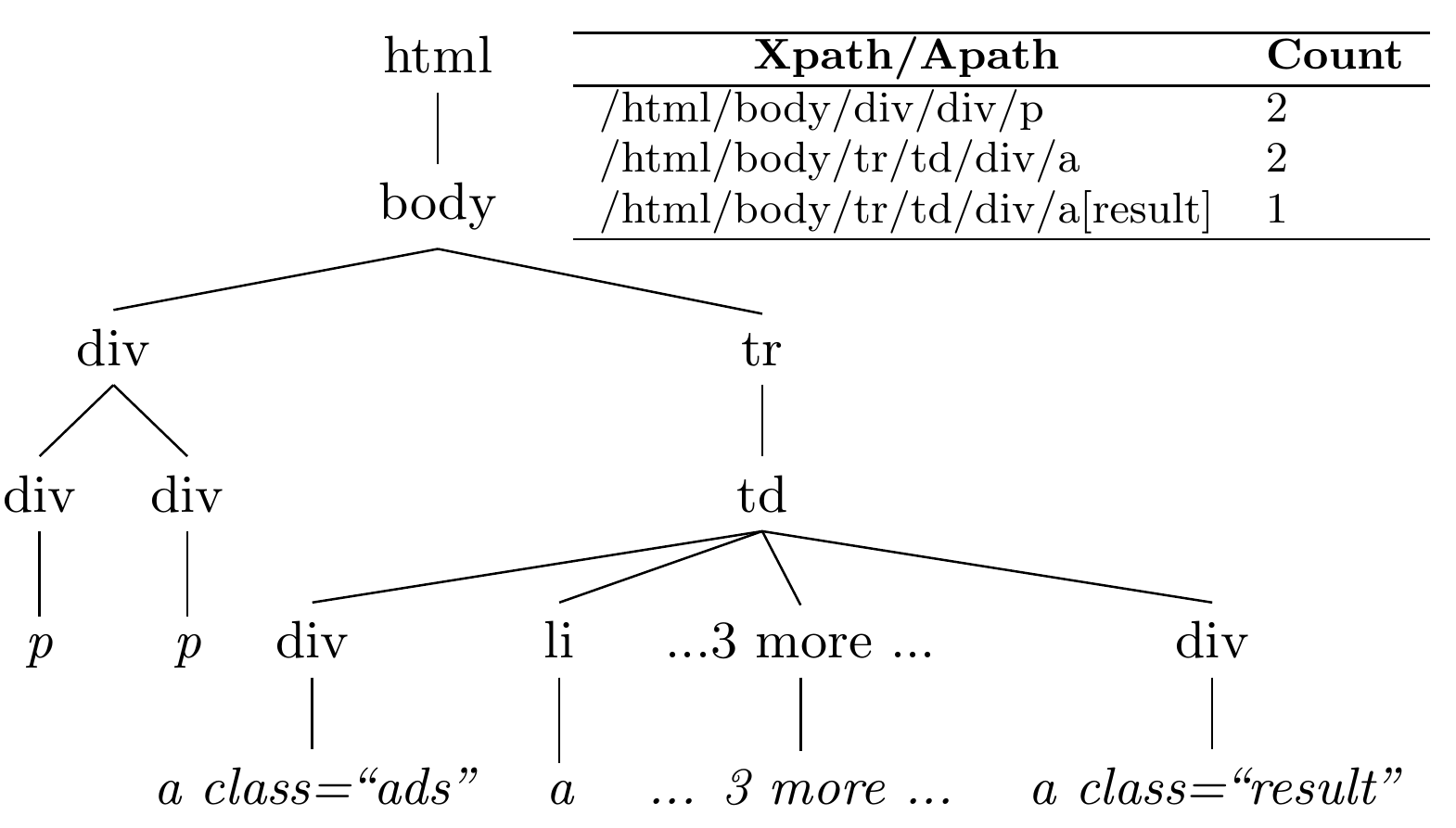}
\caption{An example of an HTML DOM Tree. Paths that end with element `a' are instances of anchor paths. [] in the Apath contains the class attributes for element `a'.}
\label{fig:dom_tree}
\end{figure}

\textbf{Anchor Path:}
Anchor links are identified by Xpaths that end with html element `a'. We call them \textit{anchor paths} (\textbf{Apath}). Note that we only focus on anchor links that direct to pages \textbf{within the same website}. 
\texttt{SOUrCe} uses all Xpaths in clustering pages and utilizes Apaths in helping crawlers to traverse the web graph. Figure~\ref{fig:dom_tree} presents an example DOM tree and several instances of Xpaths and Apaths generated from the tree. 

The same Apath may represent different semantics on the same page, especially in a table element. For example, in Figure \ref{fig:dom_tree}, the Apath
/html/body/tr/td/div/a
represents both links for advertisements and results. To reduce these ambiguities, we take advantage of Cascading Style Sheets (CSS), which describes properties of HTML elements, by adding \textbf{class} attributes for leaf elements of Apaths. 

\begin{figure}
\centering
\includegraphics[scale=0.33]{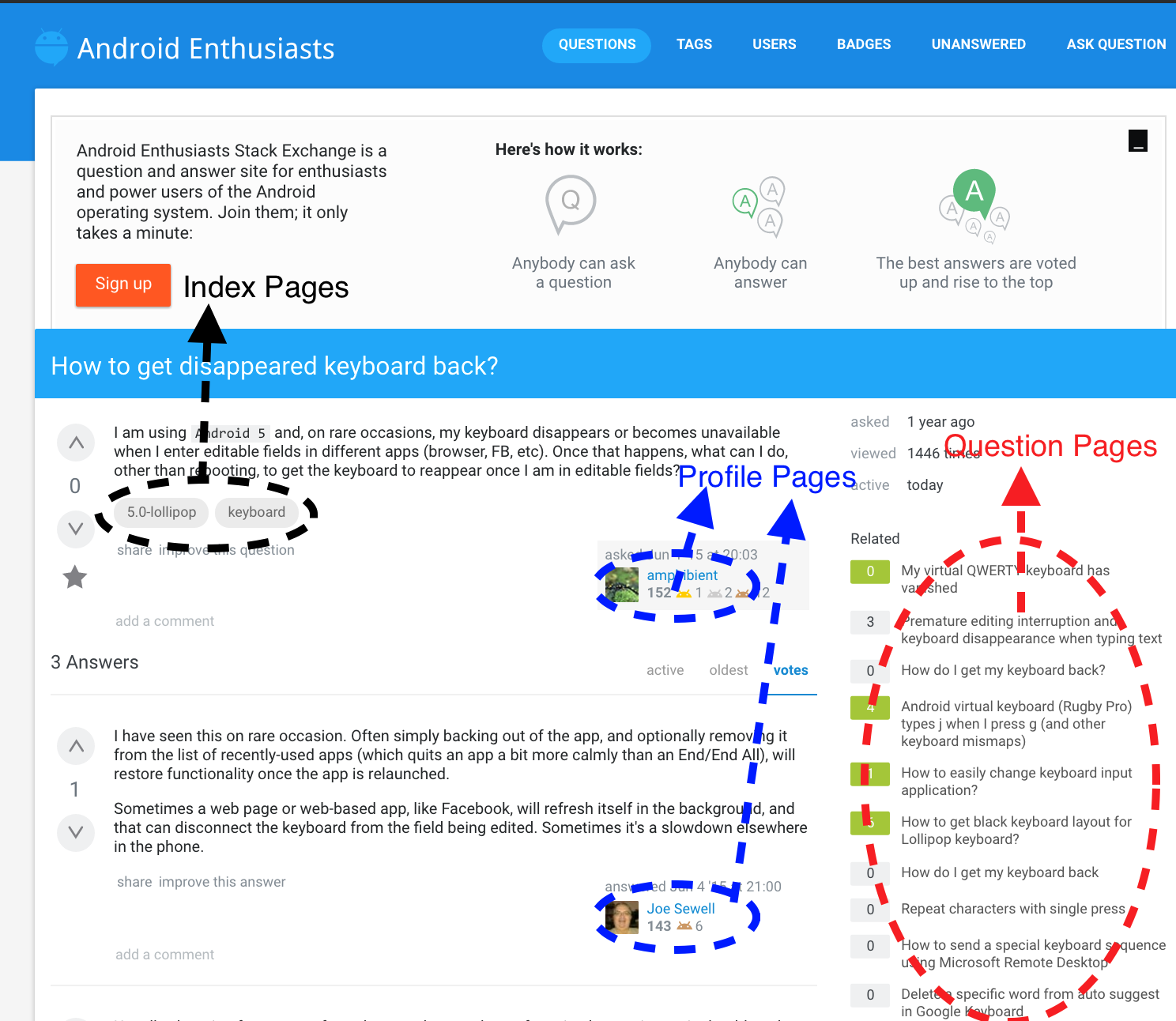}
\caption{Example of a page with identical Apaths leading to the same destination page cluster. }
\label{fig:page1}
\end{figure}

\textbf{Apath Navigation Table}: An ideal social media crawler predicts whether an anchor link is worth following before downloading its destination page. 
Unlike prior studies that utilize URL patterns and rule-based methods, \texttt{SOUrCe} generates a navigation table to guide the crawler using only structural information.~As in prior work~\cite{crescenzi2005clustering,grigalis2014using}, our intuition is that links extracted from identical Apaths on the same page are likely to point to the same destination page types. \figref{fig:page1} shows an example in which links in dashed ovals with the same color are extracted from the same Apaths and direct to the same page type. In our work, we hold a weaker assumption that the destination pages of links extracted by the same $(cluster, Apath)$ pair have the same probability distribution. Suppose we have a page $p$ from cluster $C(p)$, and extract an anchor link $u$ from $p$. Then the conditional probability distribution for the destination page $d(u)$ satisfies:
\begin{align}
\sum_{C_i \in \mathcal{C}}{P(d(u) \in C_i| C(p),x_u)} = 1
\label{eq:trans_prob}
\end{align}
where $C_i$ is a cluster from $C$, and $x_u$ represents the Apath from which $u$ is extracted in page $p$. We also denote the probability of directing to a cluster $C_i$ as $P(C_i|C(p),x_u)$. 

\section{Crawler Architecture}
\label{sec:architecture}

\begin{figure*}
\centering
\includegraphics[scale=0.55]{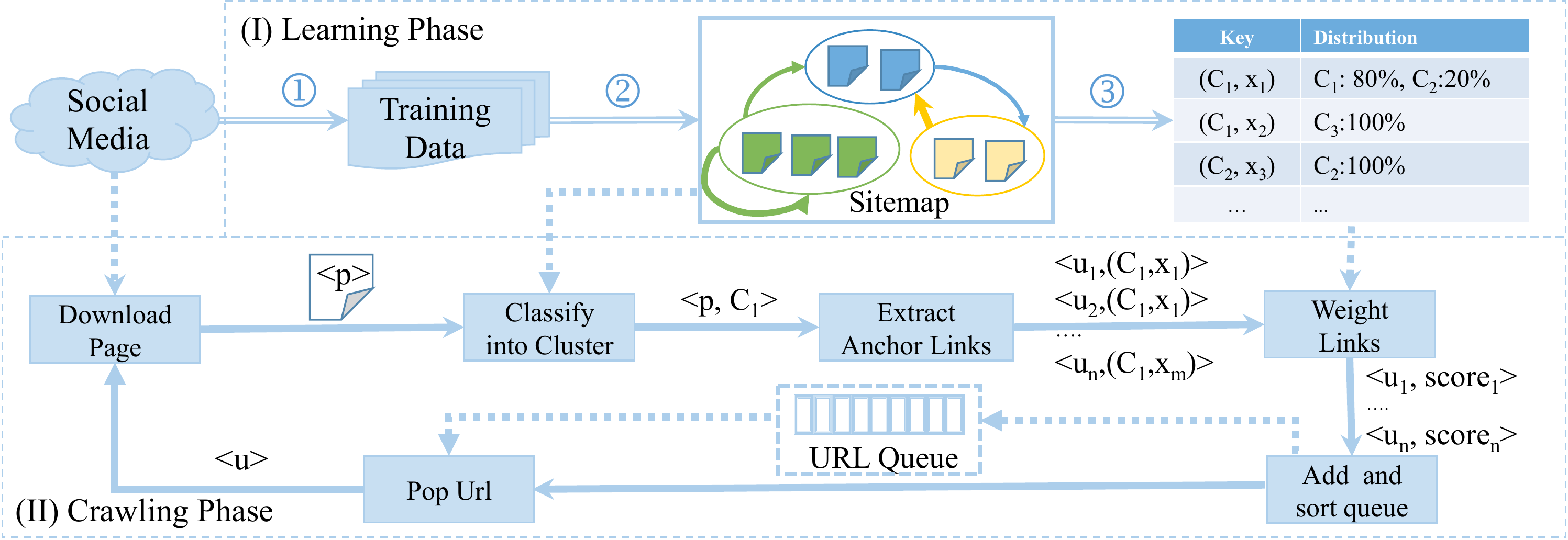}
\caption{Structure-oriented unsupervised crawler (\texttt{SOUrCe}) architecture.
Arrow $\Longrightarrow$ indicates three steps in the learning phase, concatenating three generated instances, including training data, sitemap and navigation table. Arrow $\longrightarrow$ connects flows of continuous procedures in the crawling phase and arrow $\dashrightarrow$ feeds an instance as input to a procedure.}
\label{fig:framework}
\end{figure*}

Figure \ref{fig:framework} shows the crawling architecture.  It contains two main components: (i) An unsupervised learning phase; and (ii) an online crawling phase. In this section, we introduce the learning phase in three steps, namely training data sampling, sitemap construction and  navigation table generation. Then we discuss two common crawling scenarios, crawling-for-UCC and crawling-for-target, for the online crawling phase. 

\subsection{Sampling Training Data}
\label{sec:training_data}
The first step of the unsupervised learning stage (\textcircled{1} in Figure \ref{fig:framework}) is to obtain an initial sample of pages $D$ that can be used to infer the site's structure (i.e., construct a sitemap and a navigation table). There are two requirements: (i) Most of the $(cluster, Apath)$ pairs should be covered, otherwise there will not be enough data for generating the Apath navigation table; and (ii) the training set should be as small as possible, to ensure total crawling efficiency.  In previous work, the training data were sampled using a double-ended queue~\cite{DBLP:conf/www/CaiYLWZ08} or breadth-first search (BFS)~\cite{DBLP:conf/sigir/VidalSMC06}. However, neither of these strategies are well-designed to meet these two requirements.

To explain the intuition of our sampling method, we use the example of links extracted from the red oval in Figure \ref{fig:page1}. Using breadth-first search, all of the ten links would be crawled as training data. However, since their destination pages are very likely to be in one common page cluster, usually it is not productive to crawl all of them. Thus, only one randomly-selected link is crawled for each unique Apath; other uncrawled URLs are saved in a list $U_{p,x}$ for learning the navigation table later. This strategy only uses 10\% of the bandwidth required by BFS to sample the red oval of Figure \ref{fig:page1}.  The remaining bandwidth can be used to sample other Apaths, or can be left unused to improve crawler efficiency.

\subsection{Sitemap Construction}
The next step (Figure \ref{fig:framework}, part \textcircled{2}) is to construct the sitemap.  \texttt{SOUrCe} extracts features from the training pages, as described in Section~\ref{sec:structural_feature}, and then uses DBSCAN~\cite{ester1996density} to cluster them. DBSCAN has well-defined processes for generating new clusters and dealing with outliers. Outliers represent page templates with few instances; typically they are administrative and other pages that are not the main content of the site. Usually social media is crawled to obtain the site's main areas of user-created content (UCC), which appears in pages generated from a small number of commonly-used page templates.  Outlier recognition enables trivial parts of the site to be ignored during crawling, if desired.


\textbf{DBSCAN clustering}
is a density based clustering algorithm proposed by Ester et al.~\cite{ester1996density}. It incrementally groups points that are closely packed together, and labels points in low density regions as outliers. Two parameters, $minPts$ and $eps$ are required. $minPts$ is the minimum number of points to form a dense region, also known as a cluster. $eps$ is the maximum distance between two points that are close enough to be in the same region, which is supposed to be a value specific for each dataset. In this work, $minPts$ is set to be 4, following DBSCAN authors' suggestion. 

\begin{figure}[t]
\centering
\includegraphics[width=.99\columnwidth]{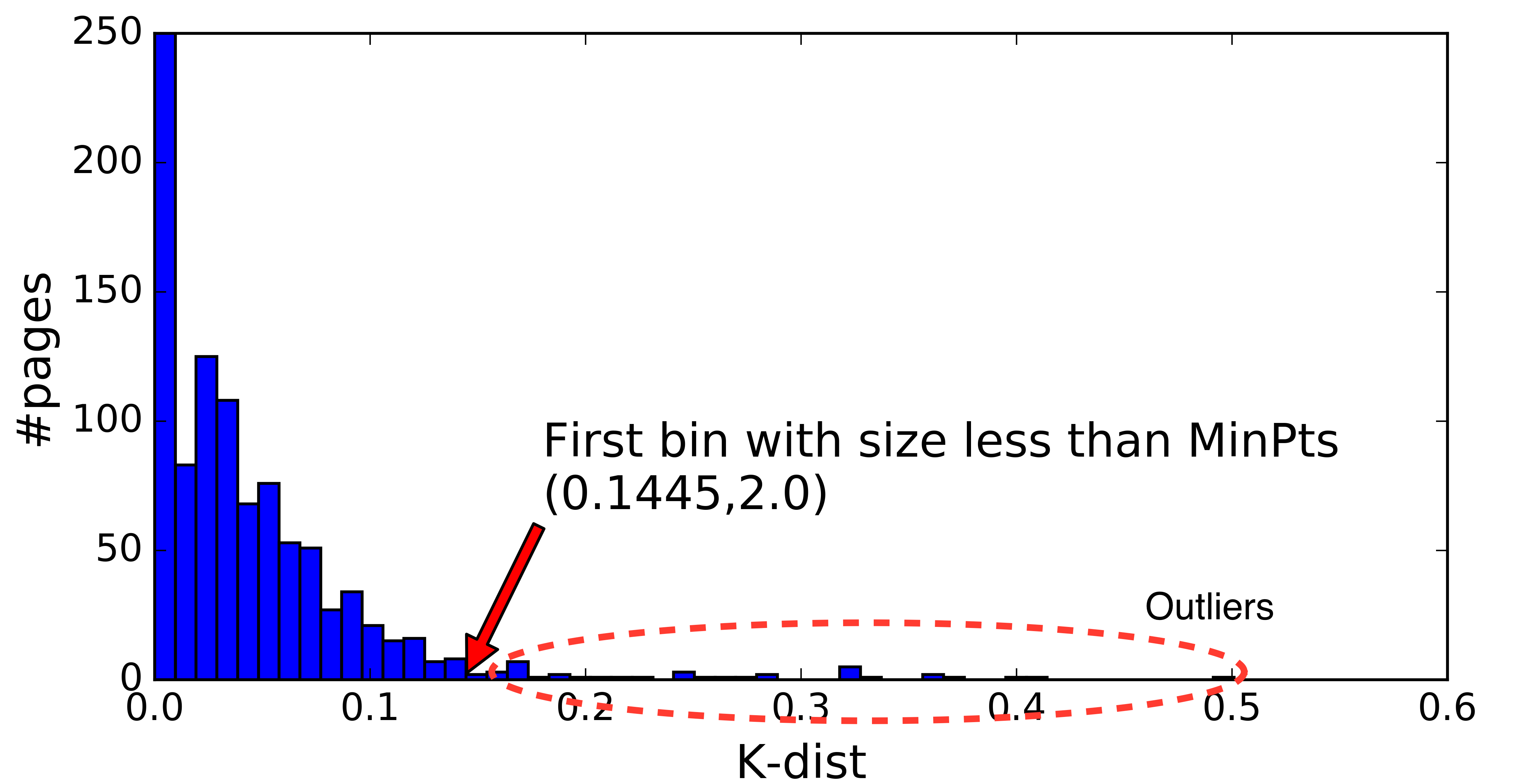}
\caption{Sorted K-dist histogram for Hupu dataset, which contains 1000 pages and 250 Xpath features. The estimated $eps$ is 0.1455 where K=3. Points with K-dist$>$0.1455 are considered as outliers. }
\label{fig:param}
\end{figure}

\textbf{Setting distance threshold}:
Previous page clustering methods usually select a constant and heuristic distance threshold for generating new clusters. However, the optimal threshold may vary for different datasets. In DBSCAN~\cite{ester1996density}, finding $eps$ is considered as finding the ``valley point'' on a K-dist graph, where $K$-dist is the distance between a node and its $K$th nearest neighbor and $K=minPts$. However, this method prefers a small value of $eps$ and creates more outliers than expected. To improve this, we plot the sorted $K$-dist histogram of the training data with $Minpts=4$. Figure \ref{fig:param} shows an example of $K$-dist histogram for the Hupu dataset. The number of bins is proportional to the number of features as: $NumBins = w \cdot NumFeats$, where $w$ is a parameter only related to the size of training data. For example, $w=4.8$ when the training data has 1,000 pages. 



From small to large $K$-dist, we find the sizes of bins typically decreases, similar to a power-law distribution. Therefore, our method chooses to set a cut-off size value to $minPts$ for bins in histogram. Specifically, it scans $K$-dist values on X-axis from small to large and stop at the first bin with size less than $minPts$ and select the X-axis value to be $eps$. Another trick to prevent small $eps$ is to ensure the cumulative number of pages at optimal $eps$ is larger than 50\% of the size of training data at the same time. An example is shown in Figure \ref{fig:param}.

\textbf{Classify a new page:} Given a newly-crawled page, its bag-of-Xpaths features are extracted, and it is classified into an existing cluster using k-Nearest Neighbor with $k=minPts-1$. Note that new pages can also be classified as outliers.  

\subsection{Apath Navigation Table Generation} 
\label{sec:navigation_table}
The third and last phase of the unsupervised learning phase (Figure \ref{fig:framework}, part \textcircled{3}) is to generate the Apath navigation table. As stated in Section \ref{sec:training_data}, during the sampling \texttt{SOUrCe} only enqueues one URL $u$ from all of the URLs extracted from the same Apath $x$ within one page $p$; but, it saves the whole URL list $U_{p,x}$ for learning the navigation table. After the sitemap is constructed, cluster labels for all pages in the training data are available.  \texttt{SOUrCe} expands URL lists with key $(page,Apath)$ to key $(cluster,Apath)$ by concatenating URL lists from pages in the same cluster:
\begin{align}
U_{C_i,x} = \bigcup_{p \in C_i;\ C_i \in \mathcal{C}}{U_{p,x}}
\end{align}
where $\bigcup$ represents the concatenation operation for URL lists. 

It is guaranteed that at least one page with a cluster label exists in $U_{C_i,x}$. Those pages with cluster labels are used to estimate the conditional transition probability of a given $(cluster,Apath)$ pair $(C_i,x)$ as follows:
\begin{align} 
{P(C_j| C_i,x)} = \frac{|d(u) \in C_j \text{ and } u \in U_{C_i,x}|}{\sum_{C_k \in C}{|d(u) \in C_k \text{ and } u \in U_{C_i,x}|}}
\end{align}
where $d(u) \in C_j$ indicates the destination page of URL $u$ is in the training data and clustered to $C_j$ in the sitemap. The adjacency matrix $A$ at the cluster level is estimated as:
\begin{align}
A_{ij} = \sum_{x\in C_i}{{P(C_j| C_i,x)} \cdot |U_{C_i,x}|}
\end{align}

\subsection{Online Crawling Scenarios}
We consider two common crawling scenarios, crawling for user-created content (\textit{crawling-for-UCC}) and crawling for a target page type (\textit{crawling-for-target}) to show how a structure-oriented sitemap supports different social media crawling strategies.

\subsubsection{Crawling-for-UCC} 

Typically when crawling social media, some types of pages are considered less interesting or useful. For example, near-duplicates and non-user-created pages are less valuable than pages shared by users. The goal is to guide the crawler to only follow links that lead to pages with user-created contents (UCC) as their main components. To achieve this goal, we propose a weighting scheme for an uncrawled URL $u$ extracted from page $p$ based on three terms, namely informativeness score, similarity score and balance score. 

$\bm{Info(C_i)}$: We first follow the idea of HITS \cite{kleinberg1999authoritative}, to calculate the authority and hub scores for clusters in sitemap with the help of adjacency matrix $A$ as follows:
\begin{align}
\label{eq:hits}
h^k &= A\ a^{(k-1)} \\
a^k &= A^T\ h^k
\label{eq:autho}
\end{align}
where hub score $h^k = (h_1^k, \cdots, h_n^k)^T$ and $h_i^k$ indicates the hub score for cluster $C_i$ at the $k$th iteration. The authority score $a^k$ is defined similarly as $h^k$.
We iteratively compute hub and authority scores until convergence. Then we define the informativeness of a cluster $C_i$ as follows:
\begin{align}
\label{eq:s}
{Info(C_i)} = \alpha \cdot a_i + (1-\alpha) \cdot h_i
\end{align}
$\bm{Dsim(C_i)}$:·
$Info(C_i)$ only depicts importance measured by link structures in the sitemap. Some constant pages can receive high authority scores but do not contain UCC, such as policy description pages that are linked by most of the pages in the site. Therefore, we introduce the term $DSim(C_i)$ to measure the structural importance of a cluster $C_i$. Clusters that contain UCC are more diverse in structures, because users are usually free to add or delete elements on those pages. In contrast, clusters that represent constant pages or duplicate pages are more likely to be similar within themselves. $DSim(C_i)$ denotes the variance of pairwise distance within one cluster, which equals to the average distance to its cluster centroid $M_i$ as follows:
\begin{align}
Dsim(C_i) = \frac{\sum_{V_j \in C_i}{\|V_j -\mathit{M_i}\|_{2}}}{|C_i|}
\end{align}
where $\|\|_2$ denotes L2 norm of a vector.

$\bm{Balance(C_i)}$: This term increases crawling diversity by penalizing too much focus on one cluster. $Balance(C_i)=1-r(C_i)$, where $r(C_i)$ is the current ratio of pages in cluster $c_i$ to total number of crawled pages. 

$\bm{Score(u)}$: The total ordering score ${Score}(C_i)$ for cluster $C_i$ is summarized as follows:
\begin{align}
\label{eq:score_c}
Score({C_i}) = Info(C_i) \cdot DSim(C_i) \cdot  Balance(C_i)
\end{align}
The score of an URL $u$ from page $p$ is estimated as follows:
\begin{align}
\label{eq:score}
Score(u) = \sum_{C_i\in \mathcal{C}} P(C_i|C(p),x_u) \cdot Score(C_i)
\end{align}
The conditional probability $P(C_i|C(p),x_u)$ can be derived from the navigation table as described in Section \ref{sec:navigation_table}, after $p$ is classified into $C(p)$. During crawling, the uncrawled URL with the highest score in the queue is selected next.

\subsubsection{Crawling-for-target}

Often it is desirable to focus crawling on pages that match the page type of an example page provided by some other process (a \textit{target} page type). For example, the goal may be to crawl all user profile pages on Quora without crawling question pages or other non-UCC pages. In this case, \texttt{SOUrCe} only focuses on crawling target pages or pages that can provide links to target pages. This scenario can be understood as a special case of crawling-for-UCC. Suppose the example page is classified into cluster $C_i$ in the sitemap. Then instead of using Equation (\ref{eq:autho}) to update authority score, we set $a_i=1$ and all other authority scores equal to 0 at each iteration. Updating hub scores still follows Equation (\ref{eq:hits}). In this case, we set $Score(C_i) = Info(C_i)$, without considering diversity and intra-cluster variance. 


\section{Experiments}

This section describes our experimental methodology and the experiments used to evaluate the \texttt{SOUrCe} crawler.

\label{sec:exp}
\subsection{Datasets}
To the best of our knowledge, there are no benchmark datasets for our task, thus we collected data from six social media sites to investigate our ideas. 
Table \ref{tab:dataset} briefly describes the datasets. Although URL patterns are sometimes not stable, we chose datasets that do have reliable URL patterns to simplify quantitative evaluation. 

\textbf{Page Type Annotation:} The ground truth labels for page types (Table \ref{tab:outlier}) were constructed in two stages.  Firstly, for each site, we randomly sampled 1,000 pages from its large collection ($>$ 20,000 pages) crawled by BFS. Then we manually clustered pages based on URL patterns, which are regular expressions. Secondly, we compared pages of different URL patterns in a browser and combined patterns that denote the same page type together. 
After combining, pages denoted by URL patterns with frequency less than $Minpts$ were labeled as annotation outliers. 

When crawling-for-target (Section \ref{sec:crawl-for-target}), the URL patterns associated with the example page provide the ground truth labels used to measure the crawler's accuracy.

\textbf{User-created Content Annotation:}
To assess the accuracy of crawling-for-UCC (Section \ref{sec:crawl-for-ucc}), we manually annotated clusters as containing user-created content (UCC) or not.
User-created content clusters have pages that are mainly content generated by users; policy description pages and login redirection pages are not UCC. 

\textbf{Path Filtering}: According to previous findings~\cite{grigalis2014using,leung2005xml}, filtering out paths with low document frequency will help to increase the accuracy in clustering documents. Therefore, we filtered out low-df Xpaths by setting a minimum frequency threshold $\theta=minPts$ which is kept constant for every dataset. The number of Xpaths after filtering for each website can be seen in Table \ref{tab:outlier}.
\begin{table}[t]
\centering
\caption{Social media sites used in our experiments. }
\label{tab:dataset}
\begin{tabular}{l|c}
\hline
  \multicolumn{1}	{c|}{Name} & Description                    \\ \hline
 \small{forums.asp.net} (\texttt{ASP})            & \small{Official Forum for ASP.NET}     \\
\small{youtube.com} (\texttt{YouTube})              & \small{Video Sharing Website}          \\ 
\small{movie.douban.com} (\texttt{Douban})         & \small{Chinese Movie Community} \\ 
\small{voice.hupu.com} (\texttt{Hupu})           & \small{Chinese Sports Forum}           \\ 
\small{rottentomatoes.com} (\texttt{Rott})       & \small{Movie Rating Community}         \\ 
\small{stackexchange.com} (\texttt{Stack})         & \small{Q\&A Website for Programming}   \\ \hline
\end{tabular}
\end{table}

\begin{table}[t]
\centering
\caption{\texttt{SOUrCe} clustering statistics for 6 datasets. $|D|=1,000$ pages per dataset.}
\label{tab:outlier}
\scalebox{0.9}{
\begin{tabular}{l|c|c|c|c|c}
\hline
Site & \multicolumn{1}{c|}{\small{Xpaths}} & \multicolumn{1}{c|}{\begin{tabular}[c]{@{}c@{}}\small{Page}\\ \small{Types}\\\small{(Annotated)}\end{tabular}} & \multicolumn{1}{c|}{\begin{tabular}[c]{@{}c@{}}\small{Page}\\ \small{Clusters}\\\small{(DBSCAN)}\end{tabular}} & \multicolumn{1}{c|}{\begin{tabular}[c]{@{}c@{}}\small{Outlier}\\\small{Pages}\\ \small{(Annotated)}\end{tabular}} & \multicolumn{1}{c}{\begin{tabular}[c]{@{}c@{}}\small{Outlier}\\\small{Pages}\\ \small{(DBSCAN)}\end{tabular}} \\ \hline
\texttt{ASP}  & 194 & 9& 12 & 5  & 13                    
\\ 
\texttt{YouTube}  & 446 & 4   &  12   &   18                                                                                    & 13                                                                                  \\ 
\texttt{Douban}  & 408 &22                           & 28                                       &       30                                                                                   & 15                                                                                  \\ 
\texttt{Hupu}  & 250 &   7                        &    11                                 &         34                                                                             &      19                                                                             \\ 
\texttt{Rott}  & 414 & 15                           & 20                                       &                                 31                                                                                   & 27                                                                                  \\ 
\texttt{Stack}  & 416 & 15                          & 20                                       &                                 24                                                                                   & 55                                                                                  \\ \hline
\end{tabular}
}
\end{table}

\subsection{Page Clustering}

Page clustering is a crucial part of the \texttt{SOUrCe} crawler. When pages are classified into the wrong clusters, the destination distribution is noisy. Therefore, we conducted two experiments to evaluate the accuracy of the clustering algorithm. The first evaluated the clustering solutions for training data $D$. The second tested classification accuracy with 4-fold stratified cross validation. 

As for baseline methods, we reimplemented method proposed by Crescenzi et al. work (\texttt{MDL})~\cite{crescenzi2005clustering} and repetitive region representation clustering in \texttt{iRobot}~\cite{DBLP:conf/www/CaiYLWZ08}.
The threshold $dt$ for \texttt{MDL} method is set to be 0.2, following authors' suggestion~\cite{crescenzi2005clustering}. Three parameters for \texttt{iRobot}~\cite{DBLP:conf/www/CaiYLWZ08} are chosen by a parameter sweep, where $Min_{Depth}=3$, $Min_{R}=0.3$ and $Min_{Epsilon}=0.5$. 


We utilize the well-known Purity and F-measure metrics to evaluate the clustering solution on training data $D$ for the clustering experiment. 
We use both micro-averaging (weighted by cluster size) and macro-averaging (equal weighting) for the F-measure, to learn about behavior on large and small clusters.



\begin{figure}[t]
\vspace{-1em}
\subfigure[Purity and F-measure on clustering training data.]
{\includegraphics[width=.99\columnwidth]{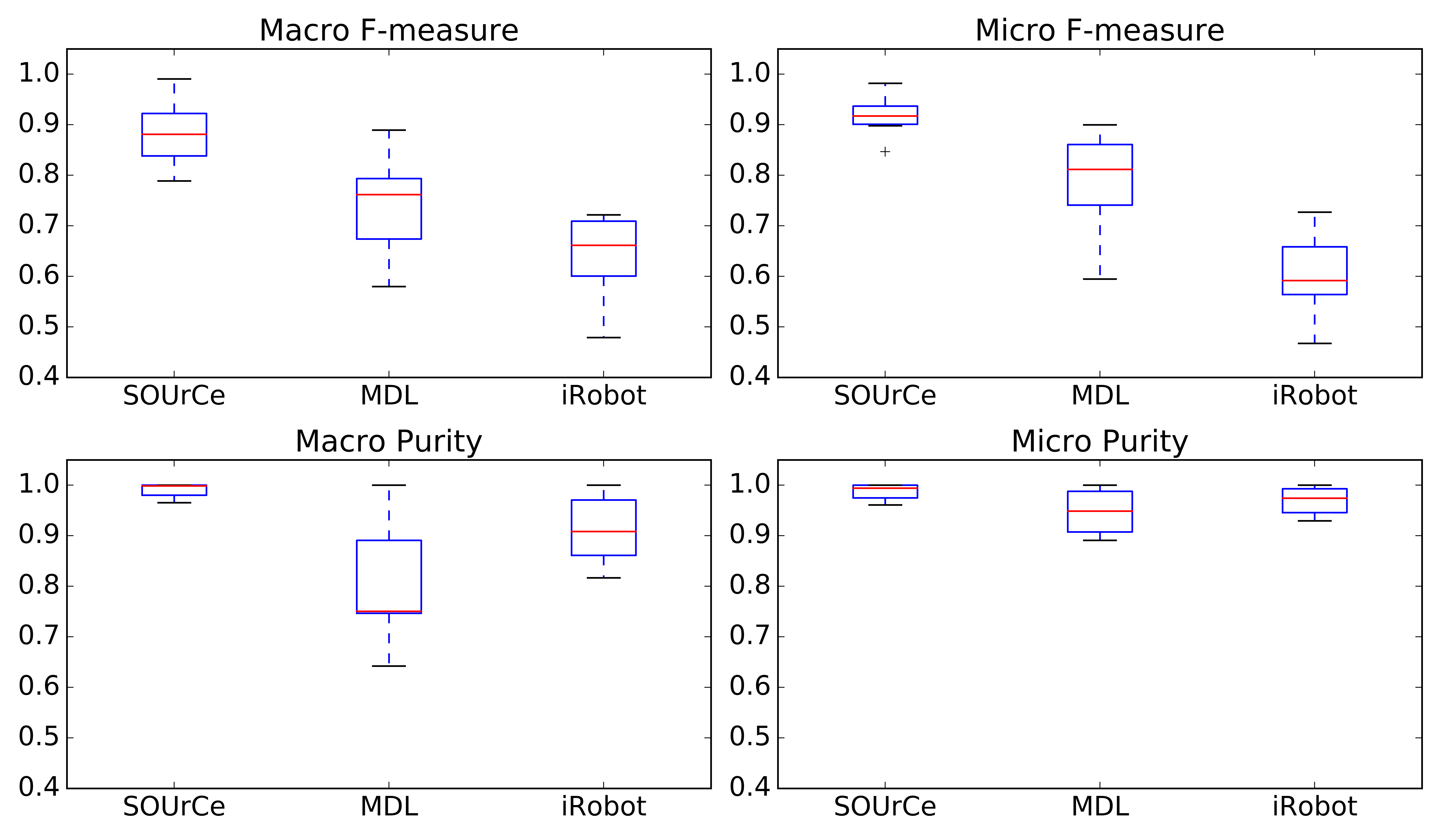}\label{fig:clustering}}
\vspace{-1 em}
\subfigure[Precision of classifying pages with 4-fold cross validation]
{\includegraphics[width=.99\columnwidth]{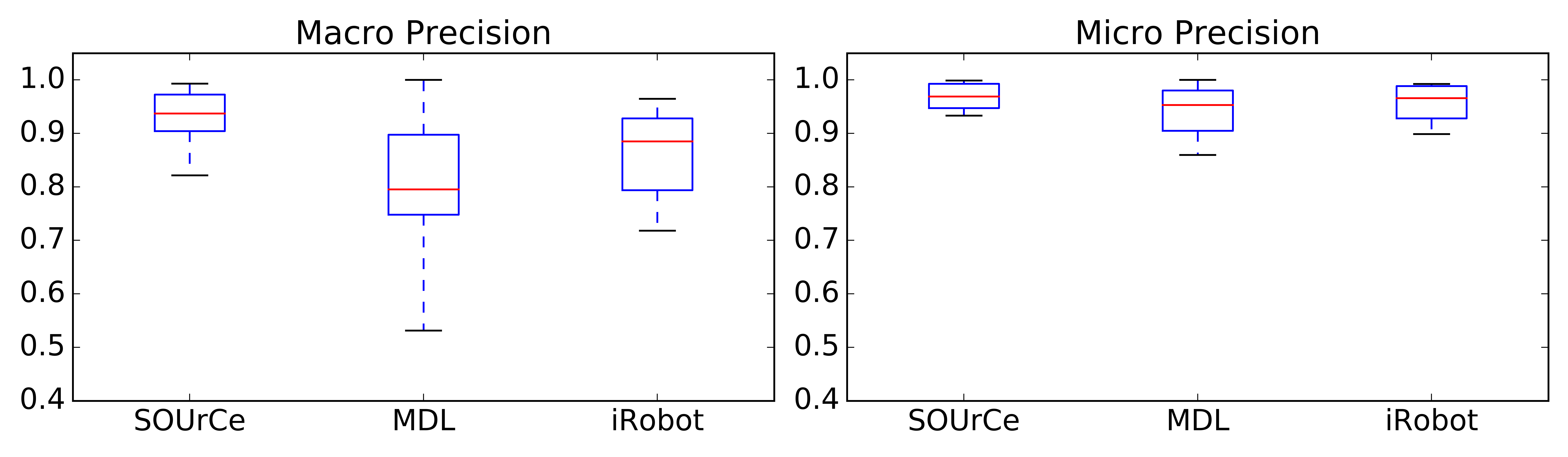}\label{fig:classifying}}
\caption{Evaluations of web pages clustering on training data $D$ over 6 datasets for two experiments.}
\label{fig:eval_cluster}
\end{figure}

As for classifying test pages into corresponding clusters, we define the Precision metric for evaluation. It first maps a page cluster $C_i$ with a page type $T_j$ and examines whether a test page is correctly classified. Different from Purity, which maps between $C^\text{test}$ and $T^\text{test}$, Precision uses training set $C^\text{train}$ and $T^\text{train}$ to do the mapping. For example, the macro- and micro- Precision are defined as:
\begin{align*}
\vspace{-1em}
\mathit{Macro\text{-}Prec}(\mathcal{C^{\text{test}}},\mathcal{T^{\text{test}}}) &= \frac{1}{|\mathcal{C^{\text{test}}}|}\sum_{i}{ \frac{|C_i^{\text{test}} \cap T_j^{\text{test}}|}{|C_i^{\text{test}}|} } \\
\mathit{Micro\text{-}Prec}(\mathcal{C^{\text{test}}},\mathcal{T^{\text{test}}})  &= \frac{\sum_{i}{{|C_i^{\text{test}} \cap T_j^{\text{test}}|}} }{N^{\text{test}}}
\vspace{-1em}
\end{align*}
where $N^\text{test}$ is the total number of pages in test data and $j=\argmax_{j}{|C_i^{\text{train}} \cap T_j^{\text{train}}|}$.


In this part, annotation outliers are not included for all methods and outliers created by clustering algorithms are considered as singular clusters. Clustering statistics for all datasets can be seen in Table \ref{tab:outlier}. 

Figures \ref{fig:clustering} and \ref{fig:classifying} show the results for clustering and 4-fold classification experiments. \texttt{SOUrCe}'s DBSCAN clustering method performs significantly better than baseline methods in both  experiments. \texttt{MDL} prefers to merge small clusters 
into an existing large one, which reduces its Purity greatly. As for \texttt{iRobot}, the features generated by repetitive regions are usually subsets of our bag-of-Xpaths model. Since the number of features is not very large, \texttt{iRobot} clustering is very likely to generate very small clusters, reducing its F-measure. Also, since the threshold parameter is constant across different datasets, \texttt{MDL} and \texttt{iRobot} have larger variances than \texttt{SOUrCe}. Macro-averaging emphasizes the accuracy on small clusters.  \texttt{SOUrCe}'s DBSCAN achieves better macro-averaged F-measure values than the two baselines.

\begin{table*}[t]
\centering
\caption{Evaluations on the crawling-for-UCC task. Best results in each metric are marked bold. }
\label{tab:crawl_info}
\scalebox{0.9}{
\begin{tabular}{c|lll|lll|lll}
\hline
 \multirow{2}{*}{\textbf{Site}}            & \multicolumn{3}{c|}{\textbf{Valid Ratio}}                                              & \multicolumn{3}{c|}{\textbf{Recall}}                                                & \multicolumn{3}{c}{\textbf{F-measure}}                                              \\ \cline{2-10}
                     & \texttt{SOUrCe}       & \texttt{iRobot}       & \texttt{BFS}          & \texttt{SOUrCe}       & \texttt{iRobot}       & \texttt{BFS}          & \texttt{SOUrCe}       & \texttt{iRobot}       & \texttt{BFS}          \\ \hline
\texttt{ASP} & 0.9988 & 0.7419 & 0.8566 & 0.3187 & 0.0466 & 0.2733 & 0.4833 & 0.0876 & 0.4144                          \\
\texttt{YouTube}            & 0.9992                  & 0.9977                  & 0.997                   & 0.2503                  & 0.0218                  & 0.2497                  & 0.4003                  & 0.0427                  & 0.3994                  \\ 
\texttt{Douban}               & 0.9996                  & 0.8994                    & 0.6064                  & 0.2907                  & 0.0437                  & 0.1764                  & 0.4504                  & 0.0833                  & 0.2732                  \\ 
\texttt{Hupu}                & 0.9978                  & 1.000                     & 0.9924                  & 0.2501                  & 0.0864                  & 0.2488                  & 0.400                   & 0.1590                 & 0.3978                  \\
\texttt{Rott}               & 0.9998                  & 0.9796               & 0.982                   & 0.2527                  & 0.0367                  & 0.2482                  & 0.4035                  & 0.07029                  & 0.3963                  \\ 
\texttt{Stack}                & 0.9984                  & 0.9979                  & 0.8328                  & 0.2978                  & 0.0277                  & 0.2484                  & 0.4588                  & 0.054                   & 0.3827                  \\ \hline
Average              & \textbf{0.9989}                  & 0.9361                  & 0.8779                  & \textbf{0.2767}                  & 0.0438                  & 0.2408                  & \textbf{0.4327}                  & 0.0828                   & 0.3773                  \\ \hline
\end{tabular}
}
\end{table*}

\begin{table*}[t]
\centering
\caption{Evaluations on the crawling-for-target task. Best results in each metric are marked bold.}
\label{tab:target}
\scalebox{0.9}{
\begin{tabular}{c|ll|lll|lll|lll}
\hline
\multirow{2}{*}{Site} & \multicolumn{1}{l|}{\multirow{2}{*}{Category}} & \multirow{2}{*}{Budget} & \multicolumn{3}{c|}{Harvest Rate}                     & \multicolumn{3}{c|}{Recall}                         & \multicolumn{3}{c|}{F-measure}                      \\ \cline{4-12} 
                     & \multicolumn{1}{l|}{} &                          & \texttt{SOUrCe} & \texttt{TPM} & \texttt{BFS} & \texttt{SOUrCe} & \texttt{TPM} & \texttt{BFS} & \texttt{SOUrCe} & \texttt{TPM} & \texttt{BFS} \\ \hline
\multirow{2}{*}{\texttt{ASP}}     & \multicolumn{1}{l|}{Detail}                    & 8314                  & 0.804             & 0.689          & 0.416          & 0.804             & 0.077          & 0.416          & 0.804             & 0.138          & 0.416          \\
\multicolumn{1}{c|}{}                         & \multicolumn{1}{l|}{Profile}                   & 4189                  & 0.846             & 0.802          & 0.265          & 0.846             & 0.090           & 0.265          & 0.846             & 0.162          & 0.265          \\ \hline
\multirow{2}{*}{\texttt{YouTube}} & \multicolumn{1}{l|}{Detail}                    & 14580                 & 0.952             & 0.994          & 0.723          & 0.952             & 0.752          & 0.723          & 0.952             & 0.856          & 0.723          \\
\multicolumn{1}{c|}{}                         & \multicolumn{1}{l|}{Profile}                   & 4244                  & 0.881             & 1.000          & 0.265          & 0.881             & 0.368          & 0.265          & 0.881             & 0.538          & 0.265          \\ \hline
\multirow{2}{*}{\texttt{Douban}}  & \multicolumn{1}{l|}{Review}                     & 1795                  & 0.772             & 0.870          & 0.096          & 0.772             & 0.101          & 0.097          & 0.772             & 0.182          & 0.097          \\
\multicolumn{1}{c|}{}                         & \multicolumn{1}{l|}{Movie}                    & 2488                  & 0.780             & 1.000          & 0.279          & 0.780             & 0.536          & 0.279          & 0.780             & 0.698          & 0.279          \\ \hline
\multirow{2}{*}{\texttt{Hupu}}    & \multicolumn{1}{l|}{Detail}                    & 15285                 & 0.971             & 0.995          & 0.816          & 0.971             & 0.136          & 0.816          & 0.971             & 0.239          & 0.816          \\
\multicolumn{1}{c|}{}                         & \multicolumn{1}{l|}{Profile}                   & 4541                  & 0.854             & 0.150          & 0.146          & 0.854             & 0.150          & 0.146          & 0.854             & 0.150          & 0.146          \\ \hline
\multirow{2}{*}{\texttt{Rott}}  & \multicolumn{1}{l|}{Detail}                    & 4722                  & 0.645             & 0.937          & 0.134          & 0.645             & 0.103          & 0.134          & 0.645             & 0.186          & 0.134          \\
\multicolumn{1}{c|}{}                         & \multicolumn{1}{l|}{Profile}                   & 8635                  & 0.816             & 0.975          & 0.534          & 0.816             & 0.425          & 0.534          & 0.816             & 0.592          & 0.534          \\ \hline
\multirow{2}{*}{\texttt{Stack}}   & \multicolumn{1}{l|}{Detail}                    & 8578                  & 0.897             & 0.997          & 0.393          & 0.897             & 0.418          & 0.393          & 0.897             & 0.589          & 0.393          \\
\multicolumn{1}{c|}{}                         & \multicolumn{1}{l|}{Profile}                   & 3854                  & 0.642             & 0.521          & 0.196          & 0.642             & 0.114          & 0.196          & 0.642             & 0.187          & 0.196          \\ \hline
\multicolumn{1}{c|}{Average}                  &                                                &                       & 0.822             & \textbf{0.828}          & 0.355          & \textbf{0.822}  & 0.273          & 0.355          & \textbf{0.822}  & 0.376          & 0.355          \\ \hline
\end{tabular}
}
\end{table*}

\subsection{Crawling-for-UCC}
\label{sec:crawl-for-ucc}

The second experiment investigated the crawler's ability to collect user-created content (UCC).  Typically user-created content is contained in the most frequent page types, which are expected to correspond to the largest clusters.

Our social media websites are too large for us to crawl and mirror entirely. Therefore we follow a previous experimental setup~\cite{DBLP:conf/www/CaiYLWZ08}, by using breadth-first search (\texttt{BFS}) to collect the first 20,000 pages of the site. Then we offer crawlers a budget $B$ to crawl at most $B<20,000$ pages within this small mirrored corpus. $B$ is set to be 5000. 
Two baselines, \texttt{BFS} and \texttt{iRobot} are adopted. We re-implemented \texttt{iRobot} in a completely automatic environment, without human supervision. Specifically, in building the traversal path for \texttt{iRobot}, we skip the step where people examine all generated paths and then remove redundant and add missing paths. Although this is not ideal for iRobot, human effort is limited in practical crawling situations, and iRobot is  compared to a fully-automatic crawler (\texttt{SOUrCe}).  For \texttt{SOUrCe}, $\alpha$ in Equation $\ref{eq:s}$ is set to be 0.5, which balances the importance of hub and authority scores, following the setting of Liu, et al.~\cite{liu2011user}.

We evaluate the quality of crawled pages using Valid Ratio and Recall, defined as follows:
\begin{align*}
    {\text{Valid Ratio (VR)}} = \frac{\text{\#(Crawled UCC)}}{\text{\#(All Crawled)}}\\
    {\text{Recall}} = \frac{\text{\#(Crawled UCC)}}{\text{\#(All UCC)}}
\end{align*}
We modify F-measure by replacing Precision with valid ratio.
We also measure the diversity of crawled results by Entropy. A higher Entropy indicates a more diverse result.
\begin{align}
\label{eq:entropy}
\text{Entropy} = - \sum_{t}{p_{t}\cdot \log_2{p_{t}}}
\end{align}
where $p_t$ is the proportion of page type $t$ in the result.


Results in Table \ref{tab:crawl_info} show that \texttt{SOUrCe} can achieve 0.9989 Valid Ratio and 0.2767 Recall on average, which significantly outperforms two baselines. High Valid Ratio indicates that the crawler is able to stay focused on user-created content.  
\texttt{SOUrCe} has a high Recall because it ranks all discovered URLs instead of filtering out unimportant URLs. 

The variance for \texttt{BFS} is very large. For \texttt{ASP}, login failure pages account for 20 percent of the corpus, however for \texttt{YouTube} the Valid Ratio is 0.997.

Although \texttt{iRobot} produces good results with human supervision, it performs less well in an automatic environment. Its overall Valid Ratio is better than \texttt{BFS} but inferior to \texttt{SOUrCe}. The reason for lower Valid Ratios on some datasets is that \texttt{iRobot} adopts a filtering method, which occasionally makes mistakes. In an automatic environment, mistakenly kept invalid nodes in the traversal path are not removed manually. Similarly, without adding missing paths to the traversal path, \texttt{iRobot}'s Recall is greatly affected. We also found that its page-flipping detection method requires a more complicated vocabulary for social media sites; simple anchor texts such as `Next', `Previous', or digits are not always sufficient.

\subsection{Crawling-for-target}
\label{sec:crawl-for-target}

The third experiment investigated the crawler's ability to focus on a particular (\textit{target}) page type, exemplified by an example page.  Detail and profile pages are usually the most common and informative types of pages~\cite{zhang2008profile}; thus, we focus on crawling these two types of web pages for all but one site. \texttt{Douban} is an exception because it requires authorization to retrieve user profiles; therefore, on this site we crawl movie and review pages instead.  

We used the same BFS-mirrored corpora used in the previous experiment. 

The total numbers of target pages in each corpus varies greatly, thus it is inappropriate to assign a fixed budget for different targets; for some (corpus, target) pairs, a fixed budget would be too generous, while for others it would be inadequate.  For example, our \texttt{YouTube} corpus had 14,580 video pages, but our \texttt{Stack} corpus had only 3,854 profile pages. Thus, we used URL patterns to find $\#(\text{All Target})$, the total number of target pages in the corpus. The crawling budget was set to be $\#(\text{All Target})$. The example target page was selected randomly from the desired page type in training data. 

$\alpha$ in Equation $\ref{eq:s}$ is set to be 0.8, because this task is focused more on finding targets than hubs. 

The baseline methods were \texttt{BFS} and Target Page Map (\texttt{TPM}) proposed by Vidal, et al.~\cite{DBLP:conf/sigir/VidalSMC06}. We modified \texttt{TPM} to support jumping to the target URL pattern on each level to make this baseline more effective. For example, for the original pattern list \{p1,p2,p3,p4\}, which only matches $p4$ in $p3$, we also allow URLs in patterns $p1$,$p2$ and $p4$ to match $p4$. This modification helps to improve the performance of \texttt{TPM}. 

In this crawling scenario, with its larger crawling budgets and tight focus on a particular type of pages, it is more likely that a crawler will reach the perimeter of our 20,000 page mirrored corpus and try to download a page that we do not have.  We do not wish to penalize a crawler for selecting a hub page that links to pages outside of the corpus.  Thus, we give this information to crawlers before crawling. For \texttt{SOUrCe}, the hub scores of those pages are preset to zero.  For \texttt{TPM}, those pages will not be selected as intermediate results to retrieve target patterns. 

We define Harvest Rate and Recall to evaluate methods on effectiveness and coverage respectively, as follows:
\begin{align*}
    \text{Harvest Rate (HR)} = \frac{\text{\#(Crawled Target)}}{\text{\#(All Crawled)}}\\
    \text{Recall} = \frac{\text{\#(Crawled Target)}}{\text{\#(All Target)}}
\end{align*}
we replace Precision with Harvest Rate to define F-measure. 

\begin{figure*}[t]

\subfigure[\texttt{Stack} Detail]{
\includegraphics[width=.67\columnwidth]{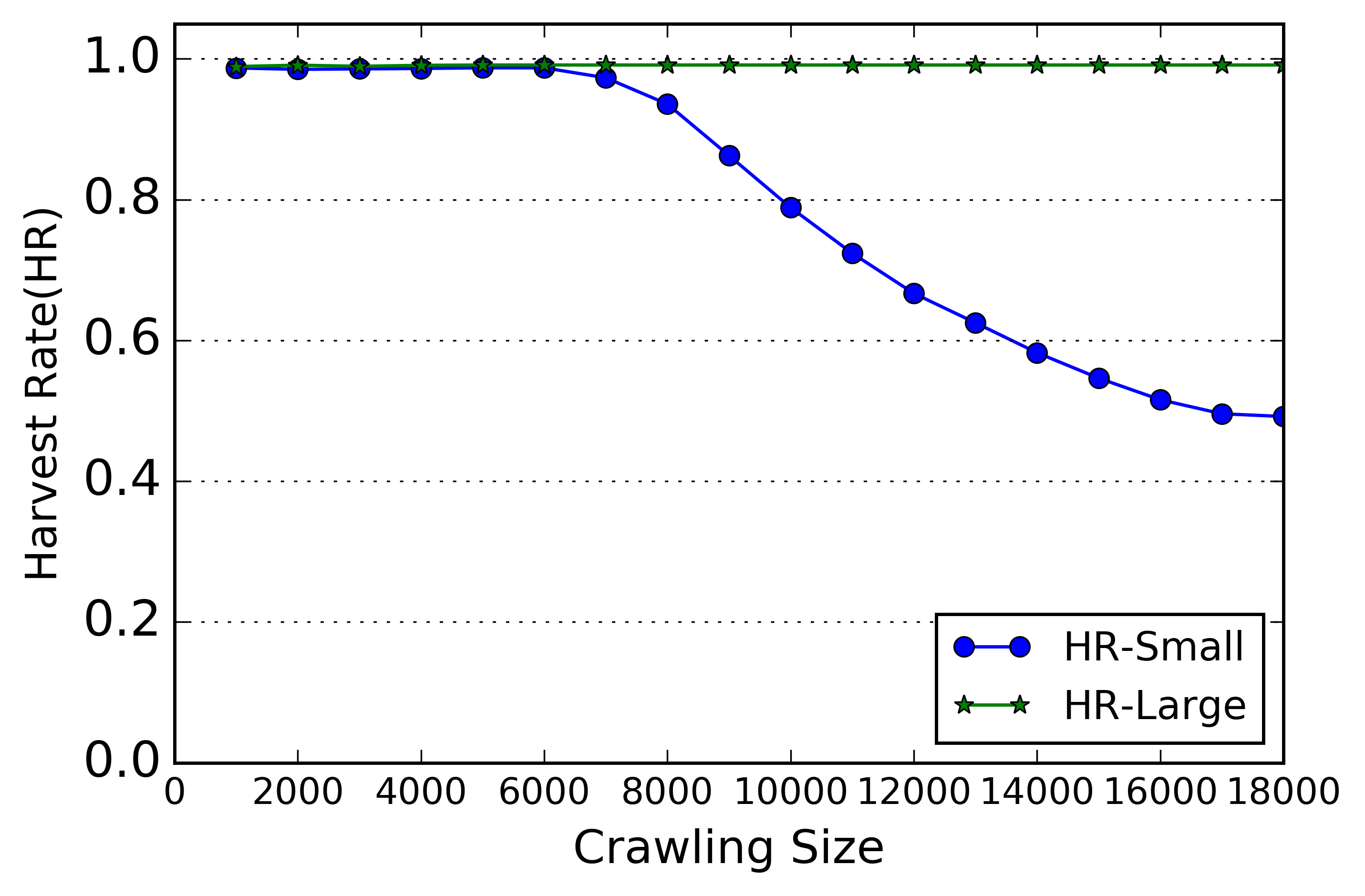}
\label{fig:stackexchange0}
}
\subfigure[\texttt{ASP} Profile]{
\includegraphics[width=.67\columnwidth]{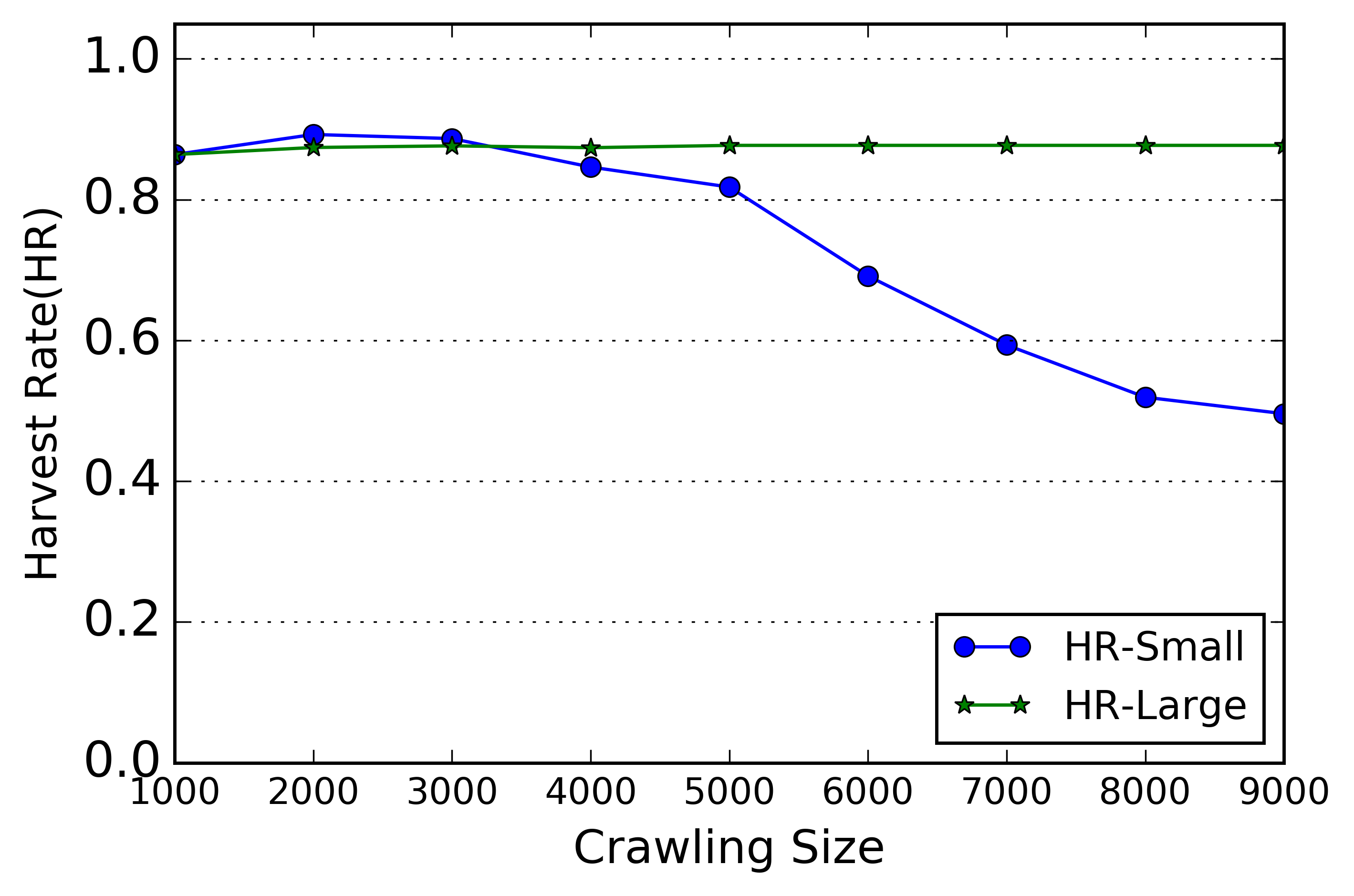}
\label{fig:asp1}
}
\subfigure[\texttt{Douban} Movie]{
\includegraphics[width=.67\columnwidth]{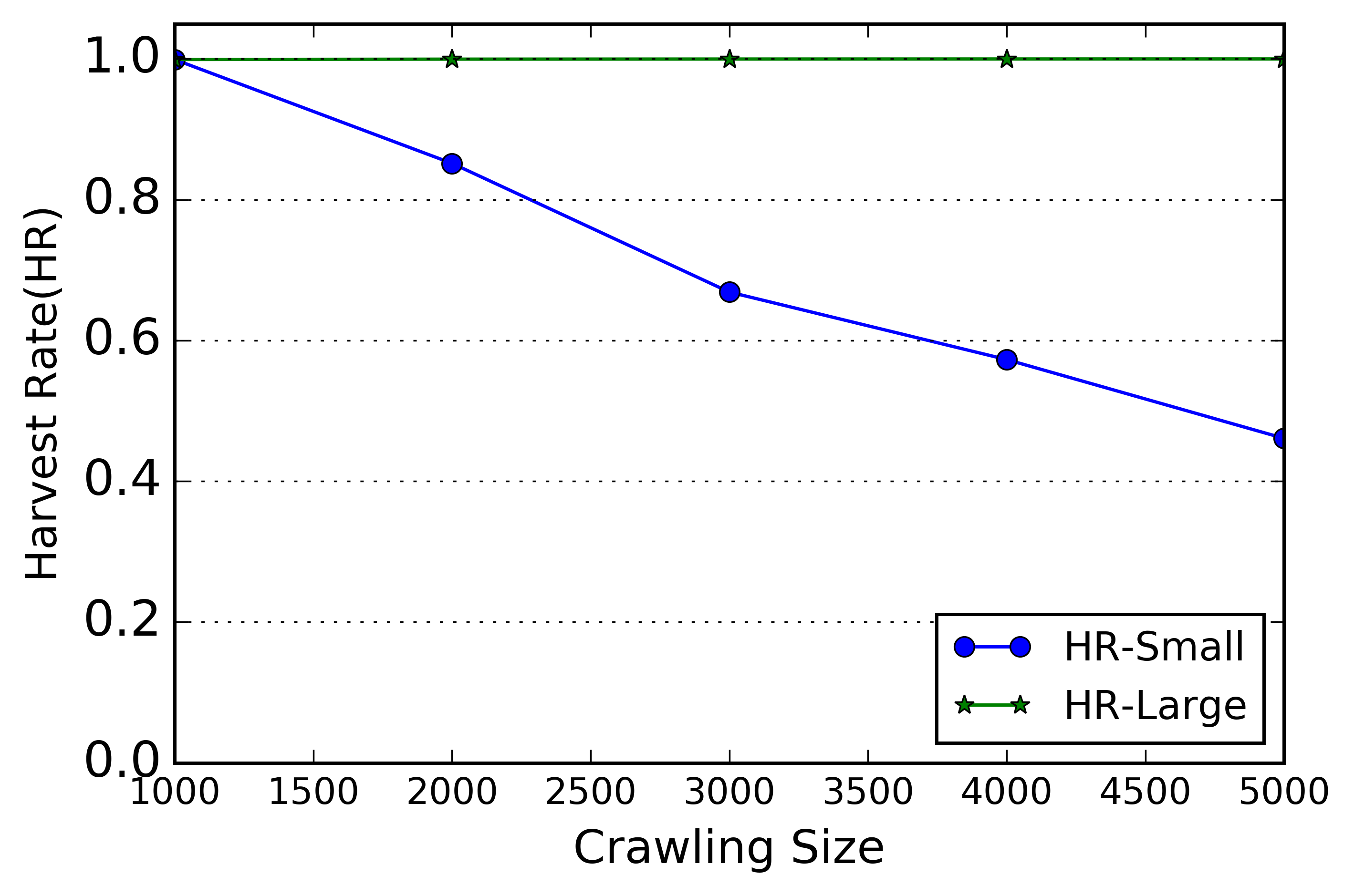}
\label{fig:douban1}
}
\vspace{-1em}
\caption{A comparison between \texttt{SOUrCe} crawling-for-target with increasing crawling size on three corpora. ``Small'' indicates the small (20,000 pages) mirrored corpora.  ``Large'' is the unbounded corpora.}
\label{fig:target_hr_examples}
\vspace{-1em}
\end{figure*}

\begin{figure*}[t]
\subfigure[\texttt{Stack} Detail]{
\includegraphics[width=.67\columnwidth]{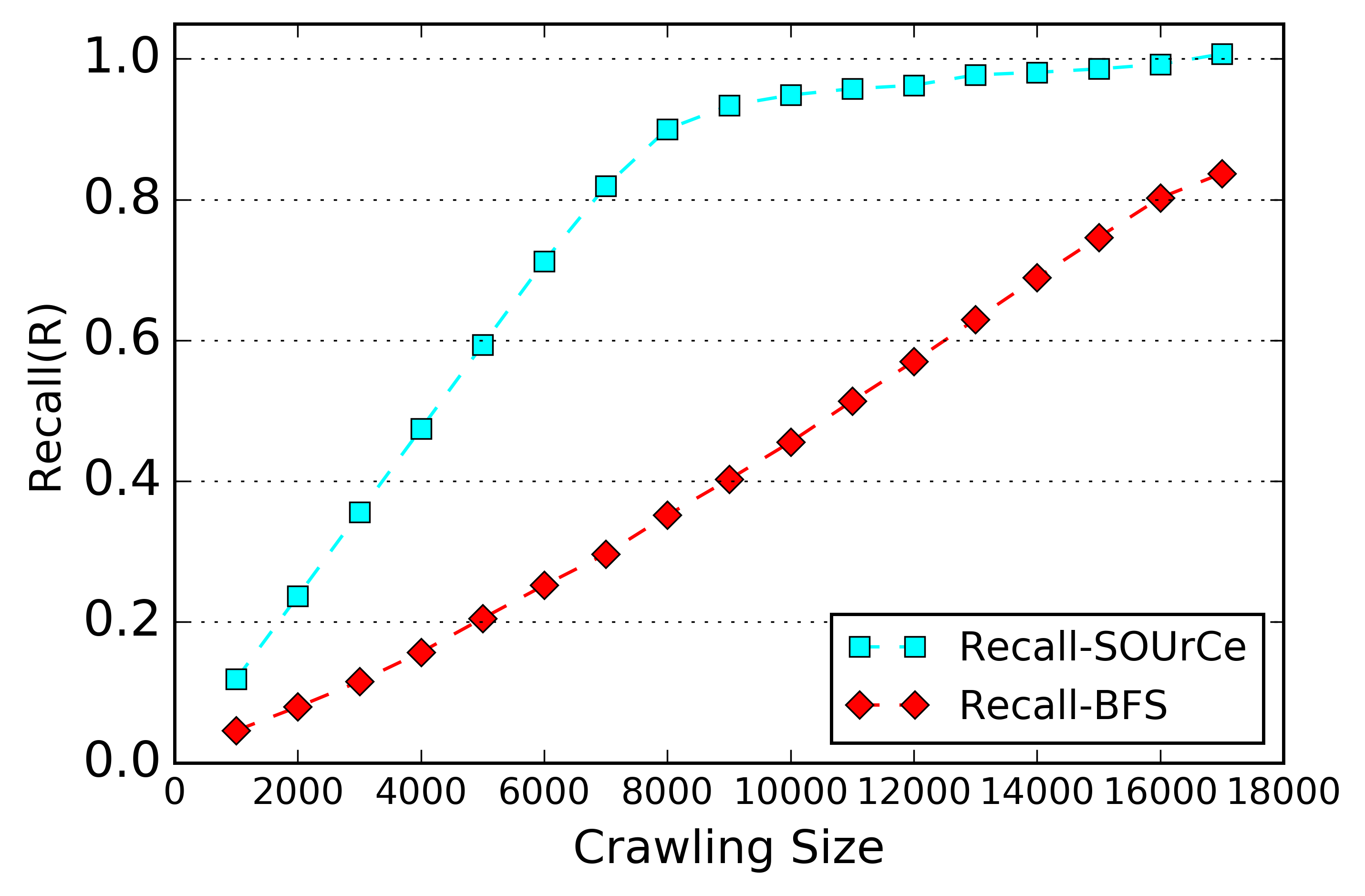}
\label{fig:r_stackexchange0}
}
\subfigure[\texttt{ASP} Profile]{
\includegraphics[width=.67\columnwidth]{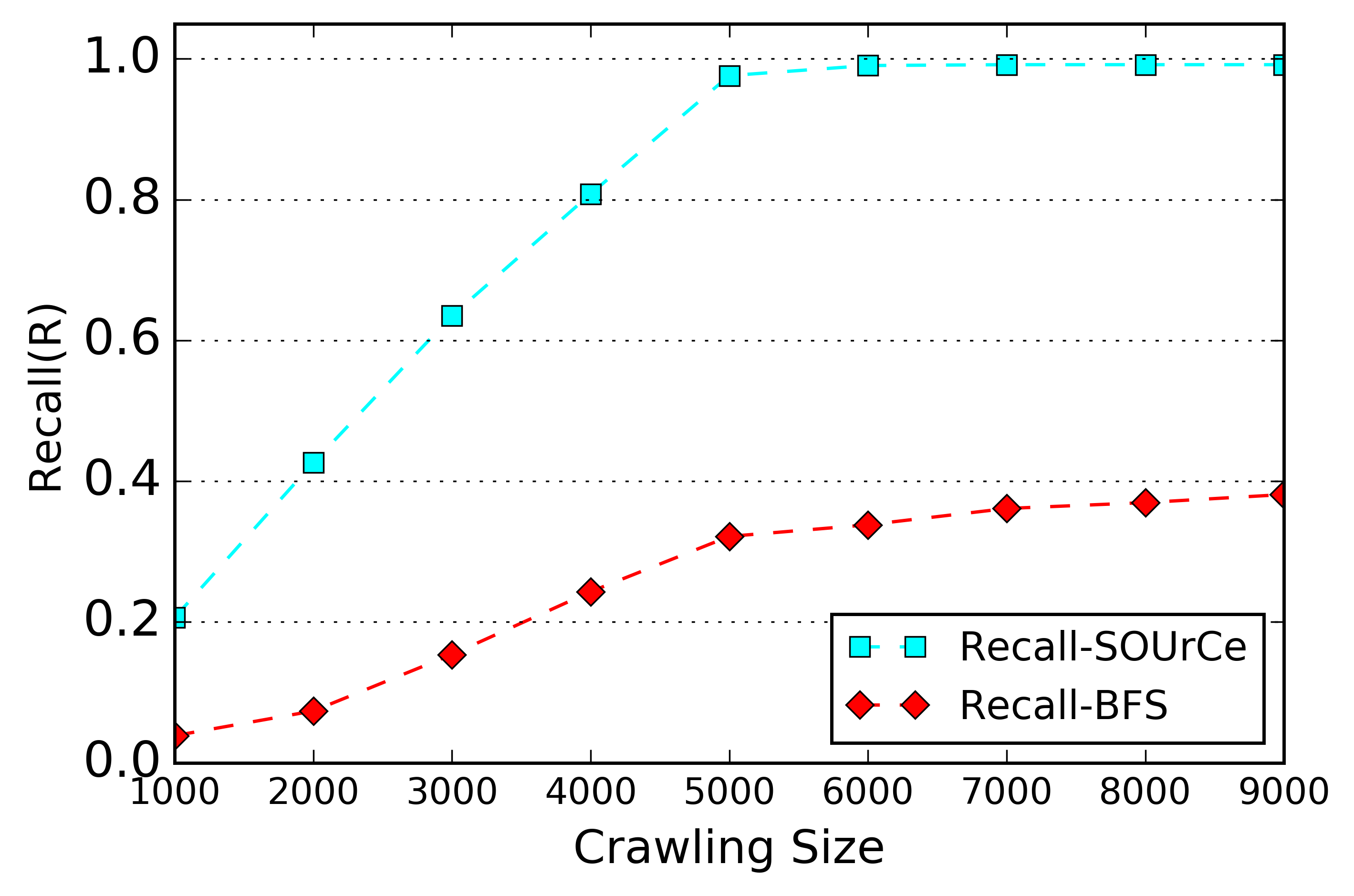}
\label{fig:r_asp1}
}
\subfigure[\texttt{Douban} Movie]{
\includegraphics[width=.67\columnwidth]{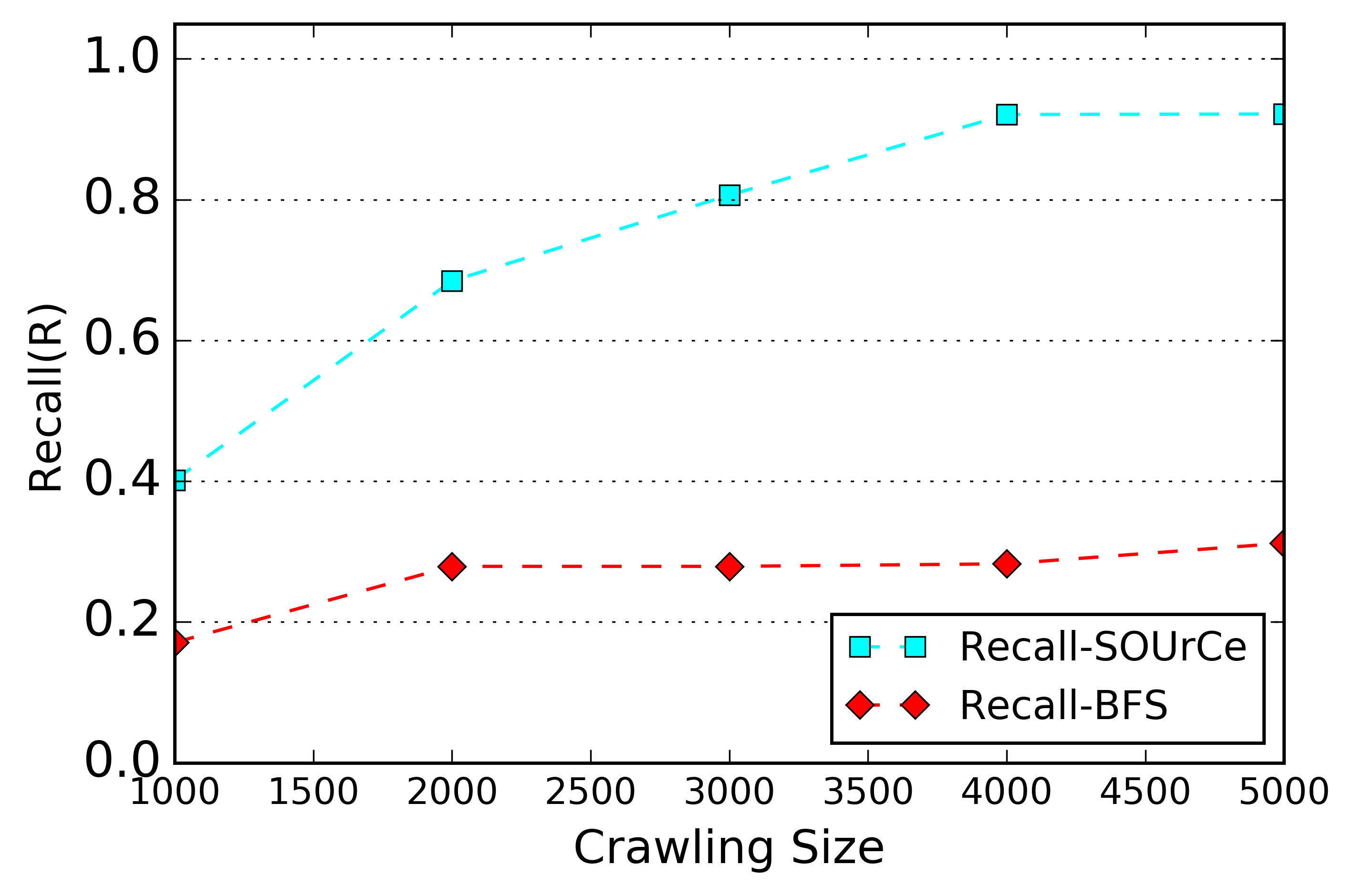}
\label{fig:r_douban1}
}
\vspace{-1em}
\caption{A comparison between \texttt{SOUrCe} and \texttt{BFS} when crawling-for-target with increasing crawl sizes on the small mirrored corpora (20,000 pages).}
\label{fig:target_r_examples}
\end{figure*}

The results in Table \ref{tab:target} show that \texttt{SOUrCe} achieves significantly better performance in Recall and F-measure than the baselines. As for Harvest Rate,  \texttt{TPM} performs similarly with \texttt{SOUrCe}. Since \texttt{TPM} is a rule-based method, it sometimes runs out of patterns and stops before reaching the crawling budget. And we also observe that \texttt{TPM} has a low harvest rate for \texttt{Hupu} Profile pages, which is caused by the failure of generating precise URL patterns for target pages automatically. 

Figure \ref{fig:target_hr_examples} shows some representative examples of the trends of \texttt{SOUrCe} with increasing crawling size. The two lines in Figure \ref{fig:target_hr_examples} represent \underline{H}arvest \underline{R}ate in Small corpus (HR-Small) and \underline{H}arvest \underline{R}ate in Large corpus (HR-Large). Here small corpus is the first 20,000 pages and large corpus is the unbounded corpus (20,000 pages, plus additional assessments as necessary when the crawler reaches an unjudged page).  The trend of HR-Small/HR-Large indicates \texttt{SOUrCe} will first select the most productive paths to follow. At the beginning of Figure \ref{fig:stackexchange0}, harvest rate equals to 1. This indicates that \texttt{SOUrCe} follows self-links, such as the related posts section or page-flipping buttons on target pages. Then we observe a drop in HR-Small, which has two causes: (1) \texttt{SOUrCe} exhausts the the most productive paths and switches to an alternative and less effective path. (2) some newly discovered target pages are already crawled previously, which makes finding new target pages difficult in late phase of crawling. The consistently high HR-Large validates the effectiveness of our method in real crawling scenarios, achieving 0.911 on average.

In Figure \ref{fig:target_r_examples}, we compare the performances of our method \texttt{SOUrCe} and \texttt{BFS} on the \textbf{Small Corpora}, evaluated by Recall. Two lines represent Recall with \texttt{BFS} (Recall-BFS) and Recall with \texttt{SOUrCe} (Recall-SOUrCe). Comparing two trends, we verify that \texttt{SOUrCe} is able to accumulate target pages at a faster rate than \texttt{BFS}. We report an average Recall-SOUrCe of 0.9498 at the final stage of crawling across 6 datasets. Figure \ref{fig:r_douban1} shows an example where the final recall for \texttt{SOUrCe} converges to 0.90 instead of 1.0. In this case, some hub pages are classified incorrectly, which caused links from those hub pages to be missed. Note that the intersect point of HR-Small and Recall-SOUrCe is where crawling size equals to $\#(\text{All Target})$. and the results of which are reported in Table \ref{tab:target}.

\subsection{Effect of Training Data}
\label{sec:size}
\begin{figure}[t]
\centering
\vspace{-1em}
\includegraphics[width=.90\columnwidth]{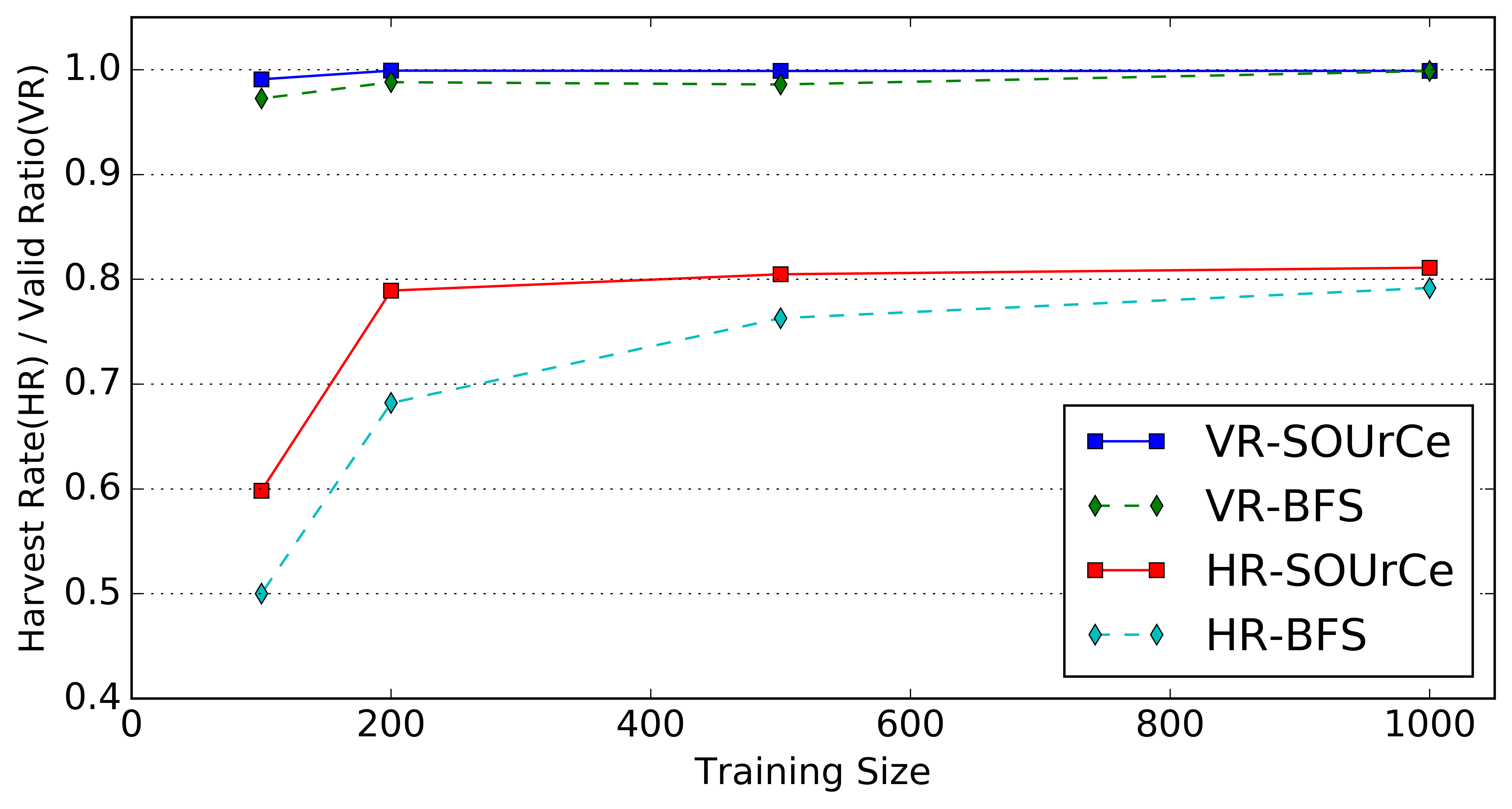}
\caption{Valid Ratio (VR) and Harvest Rate (HR) trends for different amounts of the training data $D$. Solid lines indicate our sampling method \texttt{SOUrCe} and dotted lines represent \texttt{BFS} sampling for $D$.}
\label{fig:sitemap}
\vspace{-1em}
\end{figure}

Next, we explored the influence of different amounts of training data $D$ (different sized samples of each site) and different sampling methods, for both tasks. We used the same settings and corpora as in Sections \ref{sec:crawl-for-ucc} and \ref{sec:crawl-for-target}. Two methods were utilized to sample $D$: i) the method described in Section \ref{sec:training_data}, and ii) \texttt{BFS}. With $D$ constructed by both methods, we use \texttt{SOUrCe} for crawling tasks. Results are shown in Figure \ref{fig:sitemap}. 

We find that the influence of the amount of training data is small in the range 200 to 1,000 pages.  We observe a sharp drop from 200 to 100 for the crawling-for-target task. When there is too little training data, the $K$-dist histogram for clustering is sparse, which reduces accuracy. However, the results are still good with 100 pages for the crawling-for-UCC task, which is consistent with prior research \cite{DBLP:conf/www/CaiYLWZ08}. Since the crawling-for-target task requires more precise clustering results, the performance drops faster than that of the crawl-for-UCC task. The comparisons between different sampling methods show our method is always more efficient than \texttt{BFS}. On average, 200 pages for our method performs similarly with 1,000 pages sampled by \texttt{BFS}. 

\begin{table}[]
\vspace{-1em}
\centering
\caption{Average Valid Ratio and Entropy for different combinations of features in Equation \ref{eq:score_c}. A higher Entropy score indicates a more diverse crawled result.}
\label{table:feat}
\begin{tabular}{lll}
\toprule
\textbf{Method} & \textbf{Valid Ratio} & \textbf{Entropy} \\ \hline
\texttt{Dsim+Balance}      &   0.9892    &\textbf{1.539}   \\ 
\texttt{Info+Balance}      & 0.9896       & 1.1413   \\ 
\texttt{Info+DSim}       & 0.9936      &  0.4949 \\ \hline
\texttt{Info+DSim+Balance} &    \textbf{0.9954}     & 0.8649    \\ \bottomrule
\end{tabular}
\vspace{-1.5em}
\end{table}

\subsection{Feature Importance}

We explored the importance of each term in Equation \ref{eq:score_c} by conducting an experiment in which each different term is omitted while crawling-for-UCC.  The corpus for this experiment is the entire website, instead of the first 20,000 pages, because we want to explore how each feature performs for both effectiveness and diversity in a real crawling environment. The diversity is measured by Entropy defined in Equation \ref{eq:entropy}. We used URL patterns and manual annotation to label these additional pages. 
 
 Table \ref{table:feat} shows the results. Experimental results are averaged over 6 datasets. We observe that both $Info(C_i)$ and $Dsim(C_i)$ are effective at keeping a high Valid Ratio. Incorporating $Balance(C_i)$ does increase the diversity of results. We also observe that combining $Info(C_i)$ and $Dsim(C_i)$ together reduces the diversity of results, compared with using them separately. Clusters that have high informative scores usually also have high similarity scores, such as the question pages in Q\&A sites, which have diverse structure; multiplying the two terms increases the preference towards those top clusters. Thus we observe a decrease in Entropy.

\subsection{Scalability}
\texttt{SOUrCe} has the usual crawling costs of downloading pages, parsing pages, extracting links, and maintaining a crawling queue. It also has some additional costs that other crawlers do not have; however, they are not onerous costs.

In the learning phase, the new cost involves clustering web pages represented by bags-of-Xpath vectors and generating navigation tables. The pages are obtained by breadth-first search, which is typical for undirected crawlers.  Section \ref{sec:size} shows that accurate sitemaps can be developed using hundreds of training examples.

Once learning is finished, \texttt{SOUrCe} becomes a very efficient directed crawler.  The new cost involves representing a web page as a bag-of-Xpaths, and using the sample data to perform kNN classification to order the queue.  The improved crawler efficiency far outweighs this additional cost.  Future research might consider how to prune the sample data to improve classification efficiency.

\section{Conclusion and Future Work}
\label{sec:con} 
This paper presents \texttt{SOUrCe}, which is a structure-oriented unsupervised crawler for social media sites. It consists of an unsupervised learning phase and online crawling phase for each site. In the learning phase, it efficiently samples training data and clusters pages using layout structures. Experiments show that our method can better capture structural information using the bag-of-Xpaths model and the clustering algorithm is well designed to handle noise, which provides cleaner data for generating the navigation table. The navigation table measures link structures between different clusters and is effective in sorting the crawling queue to support different types of tasks. 

The \texttt{SOUrCe} crawler architecture was evaluated under two common crawling scenarios. In tasks that aim to crawl UCC pages, the weighting scheme is very effective in measuring the importance of page clusters and achieves promising Valid Ratio and Recall. \texttt{SOUrCe} can also be modified to crawl target pages with one target example and achieves significantly better F-measure and Recall than rule-based baselines. \texttt{SOUrCe} is able to automatically detect and follow the most productive paths to find targets, especially self-links within target clusters. The high recall indicates that it finds most of the useful paths through a site. Perhaps most important, \texttt{SOUrCe} achieves its improved crawling accuracy using only small amounts of unsupervised learning that would be practical in large-scale and is an operational crawler. 

\section*{Acknowledgement}
This research was sponsored by National Science Foundation grant IIS-1450545. Any opinions, findings, conclusions or recommendations expressed in this paper are those of the author(s), and do not necessarily reflect those of the sponsor. 

\bibliographystyle{abbrv}
\bibliography{sigproc} 
\end{document}